\documentclass{INTERSPEECH2023}
\usepackage{multirow}

\interspeechcameraready

\title{\vspace{-0.1cm}Investigating model performance in language identification: beyond simple error statistics}
\name{Suzy J.~Styles$^1$, Victoria Y. H. Chua$^1$, Fei Ting Woon$^1$, Hexin Liu$^2$, Leibny Paola Garcia Perera$^3$, Sanjeev Khudanpur $^3$, Andy W.~H.~Khong$^2$, Justin Dauwels$^4$}

\address{\vspace{-0.3cm}
  $^1$Psychology, School of Social Sciences, Nanyang Technological University, Singapore\\
  $^2$School of Electrical and Electronic Engineering, Nanyang Technological University, Singapore\\
  $^3$CLSP and HLT-COE, Johns Hopkins University, USA\\
  $^4$Department of Microelectronics, Delft University of Technology, Netherlands
}

\email{suzy.styles@ntu.edu.sg, victoriachua@ntu.edu.sg, hexin.liu@ntu.edu.sg, lgarci27@jhu.edu}

\usepackage{CJKutf8}
\begin{document}

\begin{CJK}{UTF8}{gbsn}

\maketitle

\begin{abstract}
Language development experts need tools that can automatically identify languages from fluent, conversational speech, and provide reliable estimates of usage rates at the level of an individual recording. However, language identification systems are typically evaluated on metrics such as equal error rate and balanced accuracy, applied at the level of an entire speech corpus. These overview metrics do not provide information about model performance at the level of individual speakers, recordings, or units of speech with different linguistic characteristics. Overview statistics may therefore mask systematic errors in model performance for some subsets of the data, and consequently, have worse performance on data derived from some subsets of human speakers, creating a kind of algorithmic bias. In the current paper, we investigate how well a number of language identification systems perform on individual recordings and speech units with different linguistic properties in the MERLIon CCS Challenge. The Challenge dataset features accented English-Mandarin code-switched child-directed speech.
\end{abstract}

\vspace{-0.2cm}
\section{Introduction}
Linguistic variation is a fundamental characteristic of natural language. There is an increasing understanding that speech recognition and language identification~(LID) systems often underperform for groups of people who speak non-standard varieties of a language, relative to speakers of the ‘Standard’ for that language. For example, larger error rates are observed for speakers of African American English versus speakers of Californian English \cite{koenecke2020racial} and for L2 English speakers who also speak a tonal language \cite{chan22b_interspeech}. Systematic biases have also been found for individuals speaking non-Standard varieties of Estonian \cite{kukk22_interspeech} and Arabic \cite{sabty2021language}. These findings align with the growing awareness that biases in automated classification accuracy present both technical \cite{buolamwini2018gender}, and ethical \cite{birhane2021algorithmic} challenges.

Existing LID systems have shown promising results on general LID tasks. Conventional methods such as i-vector and x-vector comprise a language encoder and a back-end classifier~\cite{dehak2011language, snyder2018spoken}. Recent end-to-end LID systems integrate these two modules in a deep neural network and achieved high performance through the use of novel architectures, large data, and self-supervised learning speech representations~\cite{liu22_interspeech, efficient_lid, finetune_w2v, xlsr}.

When evaluating LID systems, system-wide metrics such as equal error rate (EER) and balanced accuracy (BAcc) are often adopted. These metrics are useful in describing the overall accuracy of a LID system (i.e., how well did the system perform in terms of identifying language spoken across an entire corpus?). In some contexts, overview statistics can be used to estimate system-level opportunity costs for different models (for example, the cost of human labour required to supplement automated processing for different datasets). However, relying solely on system-wide metrics can be poorly suited to other use cases. 

For many potential users of these systems, the end goal is to use automated metrics from individual recordings to derive secondary insights about individuals and communities. For example, in the field of language acquisition, metrics on how much an individual uses each of their languages in a given context is invaluable for understanding how multilingual skills are developing in the individual, how different combinations of linguistic input in different contexts are involved in the development of linguistic skills, and how different communities coordinate their linguistic interactions when they are with children \cite{scaff2022characterization}. Effective language use metrics are also critical in clinical contexts, where a difference in language \emph{use} may be mistaken for a difference in cognitive or linguistic \emph{ability} \cite{uljarevic2016practitioner}. At present, creating this kind of metric from human data coding is incredibly labour intensive, meaning that this kind of research is slow, difficult to fund, and creates greater barriers for research in multilingual communities – communities that are already underrepresented in language development research \cite{kidd2022diverse}. 
In this context, developmental researchers have a high need for tools that automate LID and generate estimates of language use at the level of individual recordings. Recent work has shown that LID systems underperform on accented and code-switched speech \cite{kukk22_interspeech, liu22_interspeech}. As systematic errors in such systems could underserve or disadvantage some speakers relative to others due to their language use patterns, it is important to investigate whether different models perform equally well across diverse speech samples within the corpus itself. 

In this paper, we investigate the detailed characteristics of a group of LID systems submitted to the Multilingual Everyday Recordings - Language Identification on Child-Directed Code-Switched Speech (MERLIon CCS) challenge targeted at English-Mandarin code-switched, accented, child-directed spontaneous speech. We examine how these submitted systems perform on individual recordings from speakers in a contact language environment, whose English and Mandarin language varieties differ in terms of vocabulary and pronunciation and are underrepresented in publicly available speech corpora. Moreover, as the Challenge dataset features child-directed speech, the speech register is exaggerated and different from that of adult-directed speech characteristics typically represented in speech corpora. 


The aim of this paper is to identify major difficulties that state-of-the-art LID systems present, and bridge the gap in speech research so that future models can serve end-users in related fields. In the following sections, we assess the performance of submitted systems in relation to the following variables: \emph{proportion of Mandarin spoken} (Section 3), \emph{presence of vernacular words} (Section 4), \emph{degree of child-directed speech} (Section 5), \emph{duration of language segments and rate of code-switching} (Section 6), and identify areas in which these systems collectively struggle.
\vspace{-0.4cm}
\section{Methods}
The MERLIon CCS Challenge dataset \cite{merliondata} consists of audio recordings of parent-child shared storybook reading recorded over Zoom. It is divided into Development (for system development) and Evaluation (for test) partitions. Speakers in the dataset are parents to children under the age of 5. For LID purposes, ground-truth timestamps and language labels (English or Mandarin) are provided. The speakers in the dataset use the Singaporean variety of English and Mandarin, which features different pronunciation features from other well-documented varieties of English (e.g., US and UK English) and Mandarin (e.g., Putonghua) respectively. The Challenge dataset used for this investigation is described in more detail in the Challenge description paper \cite{merlionchallenge}.

There were two tracks in the challenge: in the closed track, systems were only allowed to train on monolingual English speech curated from Librispeech \cite{librispeech} and the Singapore National Speech Corpus \cite{koh2019building}, monolingual Mandarin speech curated from Aishell \cite{aishell}, and code-switching data from the LDC SEAME corpus \cite{seame}; in the open track, systems were allowed an additional 100 hours of open or proprietary training data. Pre-trained models that were publicly available were also allowed in the open track. A total of 7 and 6 systems were submitted to the closed and open tracks of MERLIon CCS challenge respectively. As the LID task (English and Mandarin) is binary and the challenge dataset is highly imbalanced, equal error rate was selected as a primary evaluation metric \cite{duroselle2020metric}. 

Based on equal error rate, the top 3 models in the closed track were:
\begin{itemize}
\item \textbf{Multilingual ASR}: A ASR conformer model fine-tuned on Challenge Development set after removing the decoder (Team Name: Speech@SRIB).
\item \textbf{Codeswitched conformer}: A conformer model fine-tuned on code-switched utterances in LDC SEAME corpus (SBT).
\item \textbf{Ensemble 6-model}: An ensemble of 6 models fine-tuned on variable-length speech segments from Challenge Development set (Lingua Lumos).
\end{itemize} 

The top 3 open-track models were:
\begin{itemize}
\item \textbf{Whisper + Multilingual ASR}: Whisper with no fine-tuning combined with the multilingual ASR from closed track (Speech@SRIB)
\item \textbf{Fine-tuned wav2vec2.0}: A pre-trained wav2vec2.0 model fine-tuned on Librispeech phonetic attributes and Challenge Development set (UNSW Speech Processing).
\item \textbf{Ensemble titanet-1}: An ensemble of 7 titanet-1 models fine-tuned on Challenge Development Set, curated segments\footnote{Curated speech segments are 6-second segments from Mozilla Commonvoice as well as Singaporean English and Mandarin Youtube videos misidentified by ensemble 6-model in closed track.} and LDC SEAME corpus (Lingua Lumos).
\end{itemize}
In accordance to the MERLIon CCS Challenge, full model descriptions of all submitted systems can be found at the MERLIon CCS Challenge GitHub site\footnote{https://github.com/MERLIon-Challenge/merlion-ccs-2023}.

We also include two baseline models for comparison: our \textbf{MERLIon CCS baseline} conformer model trained on monolingual datasets of closed track and \textbf{Whisper baseline}, i.e., Whisper-large-v2 with no fine-tuning (industry benchmark). The submitted systems are described in more detail in the Challenge description paper \cite{merlionchallenge}. In this paper, we focus on examining differences in performance of these 8 systems on the Challenge Evaluation set.





\begin{figure}[t]
  \centering
  \includegraphics[width=\linewidth]{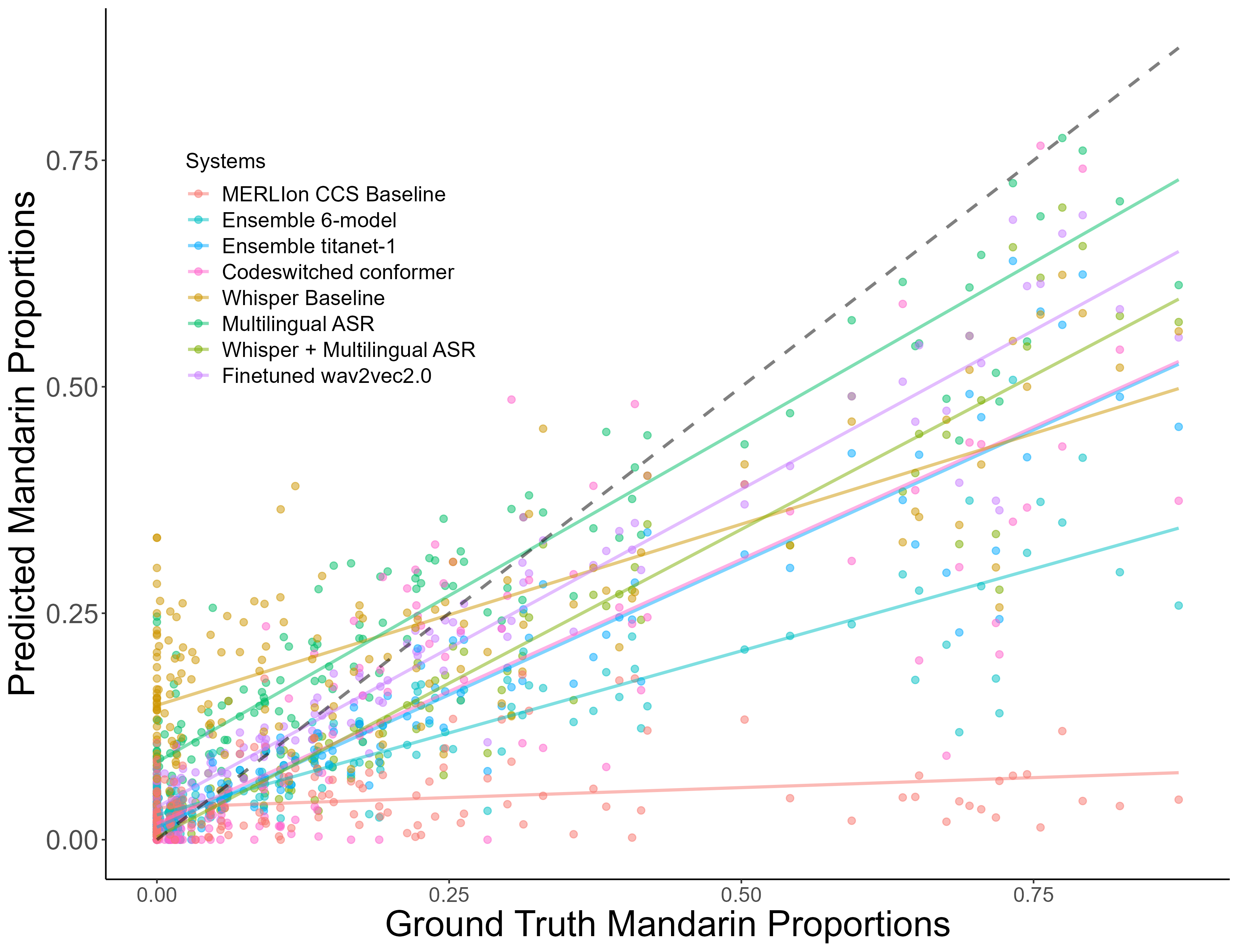}
  \caption{Proportions of Mandarin spoken per individual recording as estimated by 8 systems. Grey dotted line indicates ground-truth proportions of Mandarin spoken for each individual recording.}
  \label{fig:r_plot1}
\vspace{-0.4cm}
\end{figure}


\vspace{-0.2cm}
\section{Language Proportion Estimations}
For end-users working in language acquisition and clinical contexts, an important language identity measure is the proportion of speech in each language in an individual parent-child conversation. This means it is important to understand the accuracy of estimates in individual recordings as well as the estimate across a whole corpus. While the recordings were predominantly in English (Fig.~\ref{fig:r_plot1}), there were recordings in which speakers spoke mostly Mandarin. For each system, we map the proportions of English and Mandarin estimated against the ground-truth proportions for each recording. We present the outputs of all eight systems in Fig.~\ref{fig:r_plot1}. We observe similar trends for other systems submitted to the challenge. All models perform better on files with larger proportion of English than on files with a larger proportion of Mandarin. Performance is particularly poor when proportion of Mandarin is above 50\%. Even pre-trained systems with significant amounts of Mandarin speech data collectively underperform for individuals who speak more Mandarin.
Notably the system with the least systematic bias between conversations with high proportion of English and Mandarin, is the multilingual ASR without the pre-trained Whisper model.

 
\vspace{-0.2cm}
\section{Local Vernacular}
\begin{figure}[t]
  \centering
  \includegraphics[width=0.40\textwidth]{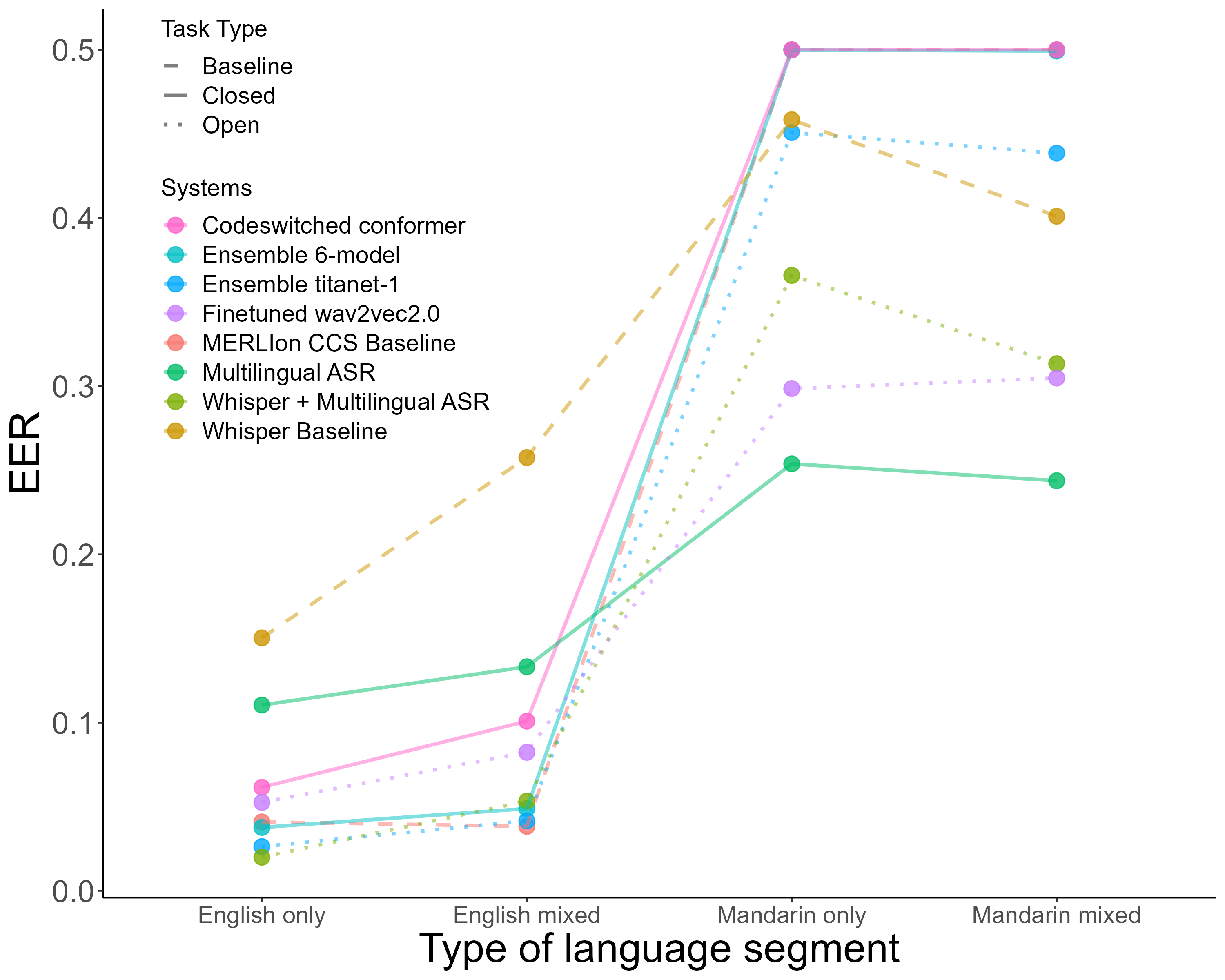}
  \caption{ EERs of each system on subsets of segments 
categorized by their language and presence of vernacular 
words. Mixed refers to segments containing English or 
Mandarin with vernacular words. }
  \label{fig:r_plot2}
\vspace{-0.4cm}
\end{figure}

The language context of Singapore can be understood as a high contact environment, resulting in lexical items, grammatical structures and accents unique to the region \cite{lee2010tonal, deterding2007singapore, sui2017spiaking}. These speech features are more pronounced in informal spontaneous speech, rarely seen in formal discourse contexts and almost never occur in corpora of read speech. Vernacular features include: clause-final discourse markers that convey mood, attitude, solidarity or emphasis \cite{leimgruber2021ethnic, smakman2013discourse}, such as `lah'; local loan words known by a large number of Singaporeans and used in speech of all languages  (e.g., `paiseh' fr. Hokkien to denote embarrassment, `makan' fr. Malay to mean eat). Many lexical items and discourse markers are historic borrowings from regional varieties including Hokkien, Teochew, Cantonese, Hakka, Tamil and Malay. These vocabulary items can appear in both English and Mandarin segments (e.g., ``don’t have \emph{lah}", ``没有 mei3you2 \emph{lah}" both of which mean ``I don't have that - DECL. EMPH."). These unique features of local vernacular speech are under-documented and under-represented in publicly available speech corpora, and may present a substantial challenge to LID systems.\\
\indent Fig.~\ref{fig:r_plot2} shows the error rates for speech segments categorized by whether they contain only lexical items from Standard Singapore English (i.e., a lexicon largely consistent with Southern British English) and Standard Mandarin, or whether they also contain local vernacular items. English segments with vernacular words tend to be misidentified more often compared to English segments without vernacular words. On the other hand, we observe mixed results for Mandarin segments, with systems tending to perform \emph{better} when local vernacular is present. \\
\indent In particular, the system with the lowest equal error rate, the Whisper pre-trained model combined with a multilingual conformer ASR model (Whisper + Multilingual ASR), shows a marked improvement in identifying “mixed” Mandarin segments compared to Mandarin-only segments. As some of the vernacular items may have been loan words in other Chinese varieties present in Singapore (e.g., Hokkien `paiseh' to denote embarrassment) or Mandarin (e.g., discourse markers like `lah' 啦 and `lor' 咯). These vernacular words may share similar phonetic attributes with the Mandarin variety represented in accessible Mandarin speech corpora (e.g., Beijing variety of Mandarin in Aishell). By contrast, although English speech is heavily represented in global speech corpora, these unique region-specific vernacular words may stand out as atypical of global English. 

\begin{figure}[t]
 \centering
 \includegraphics[width=\linewidth]{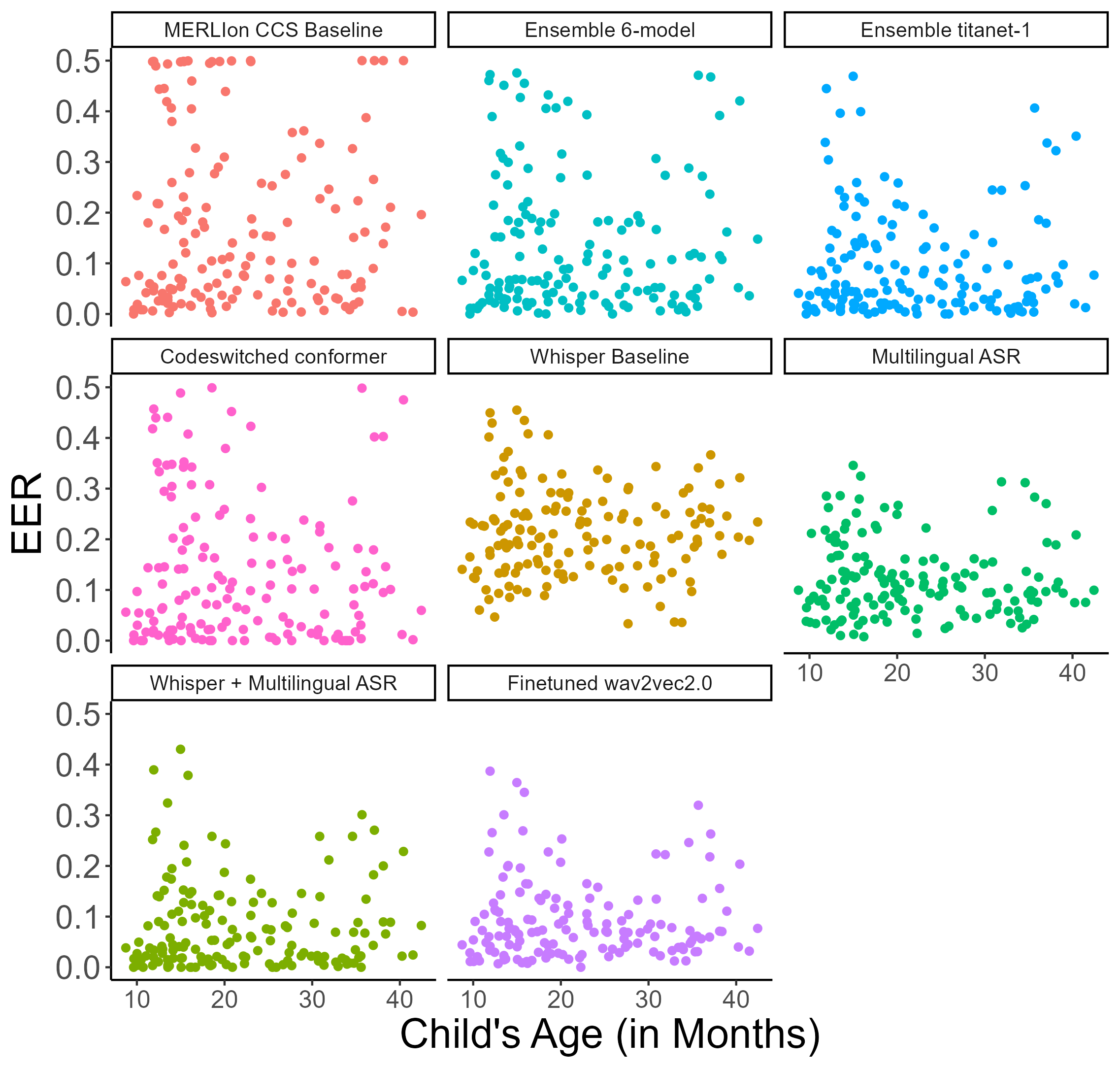}
 \caption{Scatterplots showing the relationship between equal error rates and age of child per individual recording. Each plot corresponds to a language identification system. All correlations r $<$ 0.1.}
 \label{fig:r_plot3}
\vspace{-0.4cm}
\end{figure}

\vspace{-0.2cm}
\section{Child-Directed Speech}
When communicating with young children, adults often systematically adopt a different speech style, altering the way they talk and sound. Such a register is often referred to as child-directed speech, or parentese. Child-directed speech is characterized by changes to word choice (simpler words, more repetitions), shorter syntactic structure, and unique acoustic features such as higher pitch, slower speech rate and hyperarticulation of vowel formants and tones \cite{Burnham2002}. Moreover, many features of child-directed speech also occur in other low-intelligibility speech contexts including speech produced for foreigners \cite{UTHER20072}, for artificial speech processing systems \cite{robot} and when audio clarity is reduced ~\cite{Burnham2002, hazan2015}. Recent work has demonstrated that speech recognition and LID systems underperform when speech segments have prosodic or acoustic features that deviate from the features they are typically trained on \cite{chan22b_interspeech,kukk22_interspeech}. Thus, reducing errors on speech containing these features has implications beyond child-directed speech alone. 

In the MERLIon CCS dataset \cite{merliondata}, 85\% of the speech is from parents of children under the age of 5. It has been found that as child age increases, parents employ less child-directed speech \cite{foulkes2005phonological}. We use age of child at the time of recording as a proxy for the degree of child-directed speech in the recording. As such, we may expect to see a positive relationship between LID performance and age of child. 

Contrary to what we expected, the age of child did not systematically affect the performance of LID systems, suggesting that features of child-directed speech present in our corpus do not disrupt model performance selectively for parents addressing children of different ages (Fig.~\ref{fig:r_plot3}). 
\vspace{-0.2cm}
\section{Code-Switching and Segment Lengths}

\begin{figure}[t]
  \centering
  \includegraphics[width=\linewidth]{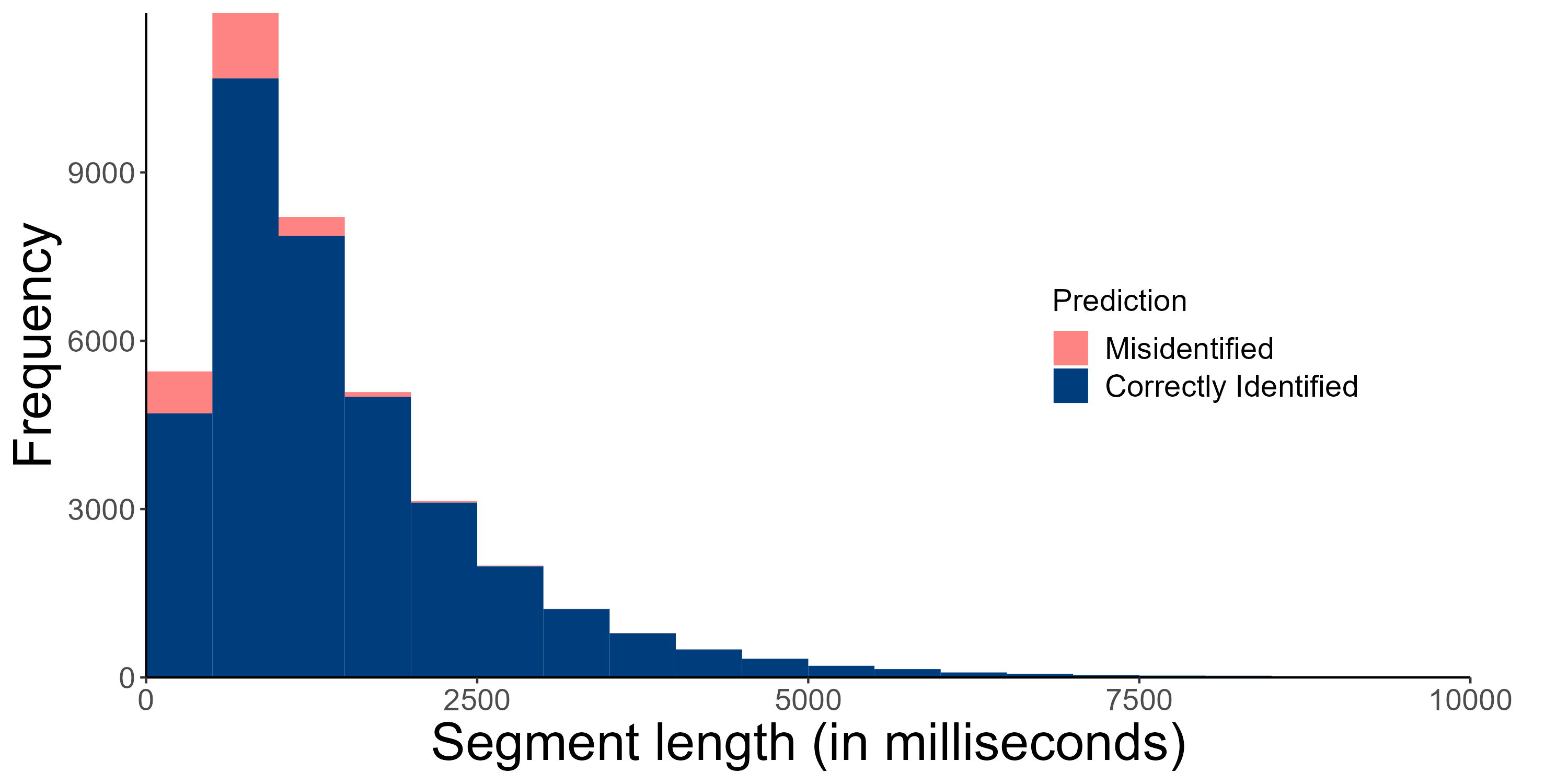}
  \includegraphics[width=\linewidth]{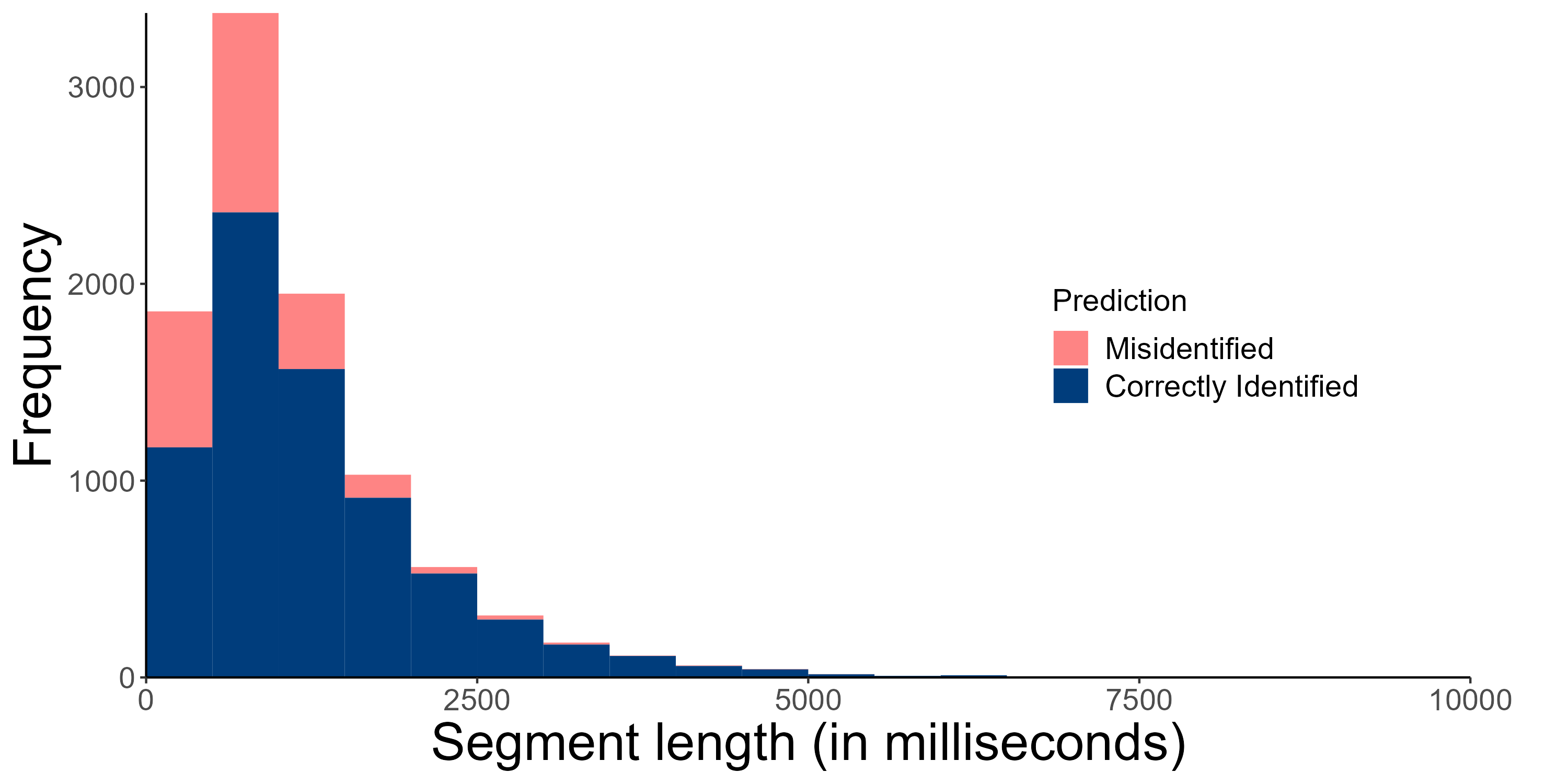}
  \caption{Frequency distributions of lengths of TOP. all English segments (N = 39245) and BOTTOM. all Mandarin segments (N = 9540) that are correctly and incorrectly identified by system with lowest EER (Whisper + Multilingual ASR).}
  \label{fig:r_plot4}
\vspace{-0.4cm}
\end{figure}

For multilingual speakers, there is increased variability and flexibility in the languages they may choose to use during spontaneous speech.  As such, the length of speech segments in each language can vary across speakers in a spontaneous code-switched speech corpora compared to a monolingual speech corpus.  Child-directed speech also has shorter sentences than adult-directed speech \cite{genovese2020infant}. As language segment length becomes shorter, it can be increasingly difficult for models to extract sufficient acoustic features for LID, e.g., the average number of unique phonemes reduces exponentially with segment duration \cite{poddar2018speaker}. Across the 8 systems, we observe that systems tend to struggle with identifying the language in segments shorter than 2 seconds, as the majority of misidentified segments lie between 0.5 to 1.5 seconds. Most misidentified English segments range from 0.5 to 1 seconds (Fig.~\ref{fig:r_plot4}), while most misidentified Mandarin segments lie between 0.5 to 2 seconds (Fig.~\ref{fig:r_plot4}). Only the performance of the winning system is shown but the trends are the same for all other systems. 

\begin{figure}[t]
 \centering
 \includegraphics[width=\linewidth]{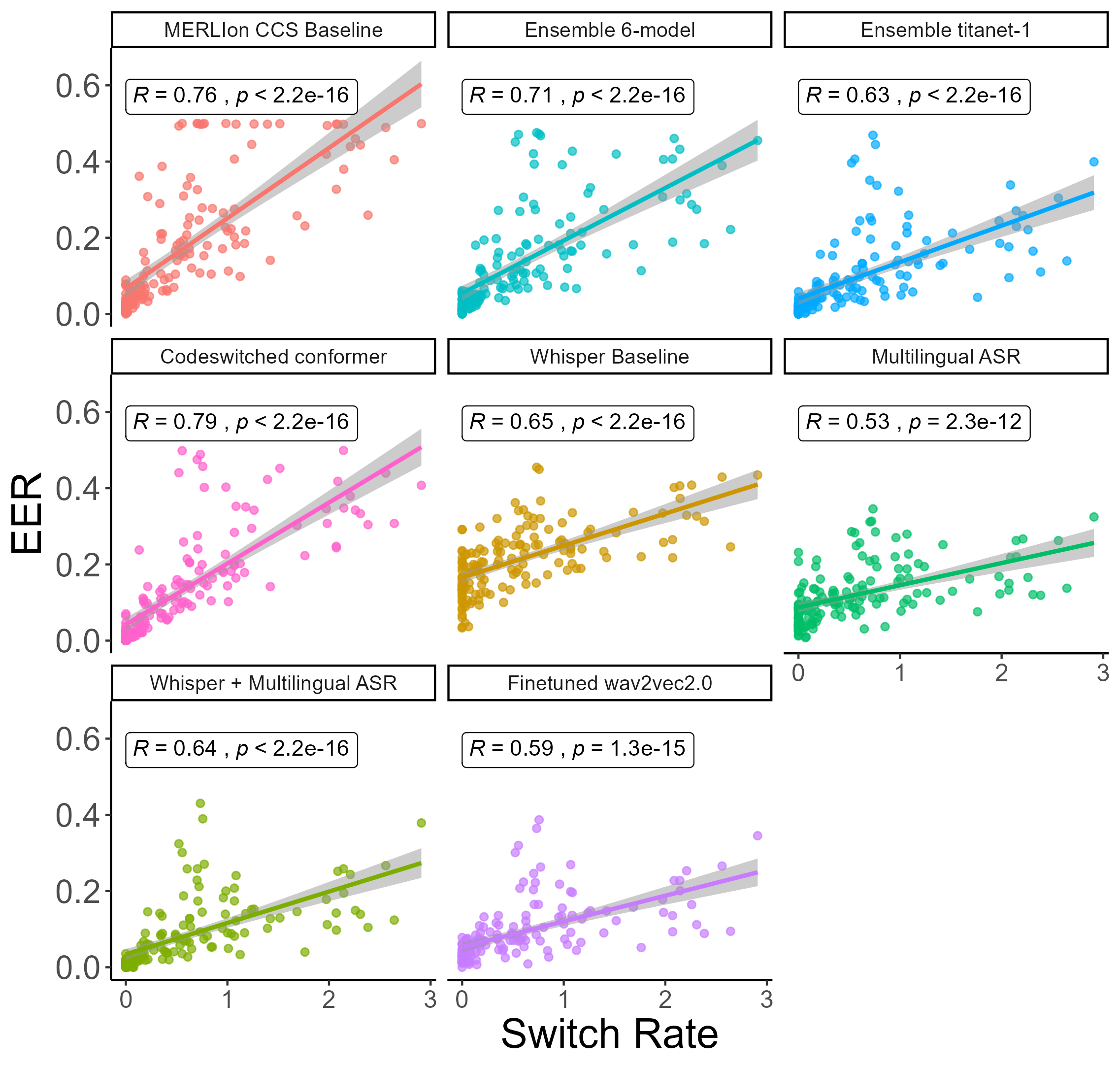}
 \caption{Scatterplots showing the relationship between equal error rates and rate of code-switching per individual recording. Each plot corresponds to a language identification system.}
 \label{fig:r_plot5}
\vspace{-0.6cm}
\end{figure}

As the speech in our dataset is spontaneous, the rate of language switching also varies across recordings. For instance, two recordings that have equal proportions of English and Mandarin may differ in the number of language switches the speaker may choose to employ, either switching back and forth between languages or only switching to another language once. As the presence of language switches is closely related to length of speech segments in each language (i.e., more switches, shorter speech segments), it is important to investigate if LID systems struggle with speakers who employ more language switches. 
To that end, we computed the rate of code-switching in each recording. Rate of code-switching per minute is defined as the number of switches to another language between each speech segment divided by total minutes of speech per recording. For instance, a speaker who speaks in English, then Mandarin, back to English again, in one minute of speech, has a rate of 2 language switches per minute. As expected,  we observe that LID systems make significantly more errors as the rate of code-switching increases (Fig.~\ref{fig:r_plot5}). 

In general, all systems struggled with LID in short segments of speech, which was compounded by high switch rates. In addition, performance was poorer overall for predicting Mandarin segments, especially when segment length was short. 

\vspace{-0.2cm}
\section{Conclusion and Future Directions}
Overall, the results of the present study reveal systematic biases in the LID systems submitted to the MERLIon CCS Challenge. There are two main findings we wish to highlight in the present work. First, our findings provide foundational evidence that spoken corpora used in the training of large pre-trained systems under-represent the range of spontaneous speech from certain groups of speakers, specifically; speakers who speak less English, switch between languages more frequently, or use more vernacular words in their English. By contrast, LID systems are unaffected by features of child-directed speech produced for children of different ages. Second, we observe that the presence of vernacular words impacts LID accuracy of English and Mandarin differently. When English segments are mixed with vernacular words, systems tend to struggle with them, while the effect is not so pronounced for Mandarin LID. These words are underrepresented in most English speech corpora, contributing to its poorer performance. Conversely, region-specific words share more acoustic commonalities with Mandarin. In a contact language environment like Singapore, we recommend language models that are targeting code-switching not only train on monolingual resources but target common lexical items that occur between language pairs or multiple language groups. 

Models that have been pre-trained on larger datasets do show performance advantages across the board, even though their pre-training datasets are not representative of local speech varieties. However, end users may wish to reduce biases by selecting a model with lower overall performance but less discrepancy in error rate across languages, at the level of individual recordings.
Taken together, we demonstrate that additional investigations of multiple metrics alongside system-wide EER across the corpus could clarify how well LID systems are performing for different groups of speakers in a corpus. We recommend the adoption of additional metrics during system evaluation to ascertain the individual variation in error rates for automatic speech processing systems. 
\vspace{-0.2cm}
\section{Acknowledgements}
This work was partially supported by the National Research Foundation, Singapore, under the Science of Learning programme (NRF2016-SOL002-011), the Centre for Research and Development in Learning (CRADLE) at Nanyang Technological University (JHU IO 90071537), and the US National Science Foundation via CCRI Award \#2120435.


\vfill\pagebreak
\bibliographystyle{IEEEtran}
\bibliography{mybib}
\end{CJK}
\end{document}